\documentclass[a4paper,dvipdfmx,nofootinbib]{revtex4}
\usepackage{graphicx}
\usepackage{fancyhdr}
\usepackage{amsmath}
\usepackage{color}

\begin{document}

%Title of paper
\title{Beyond the SM phenomena and the extended Higgs sector based 
on the SUSY gauge theory with confinement}

% Repeat the \author .. \affiliation  etc. as needed
%
% \affiliation command applies to all authors since the last
% \affiliation command. The \affiliation command should follow the
% other information

\author{T. Shindou}
\affiliation{
Department of Applied Physics, School of Advanced Engineering , Kogakuin University,
Tokyo 163-8677, JAPAN}

\begin{abstract}
We propose a fundamental theory whose low-energy effective 
theory provides a phenomenological description of electroweak baryogenesis, 
radiative neutrino mass generation, and dark matter.
The model is based on SUSY SU(2)$_H$ gauge theory with confinement, 
and the model contains new $Z_2$ discrete symmetry and $Z_2$-odd  right-handed 
neutrino superfields. The Higgs sector in the low energy 
effective theory of this model below confinement scale is described by 
fifteen mesonic superfields of fundamental SU(2)$_H$ doublets.
We present a benchmark scenario of this model, where all the constraints from  
the current neutrino, dark matter, lepton flavour violation and LHC data are satisfied. 
We also discuss how to test the scenario by the future collider experiments.
\end{abstract}

%\maketitle must follow title, authors, abstract
\maketitle

\thispagestyle{fancy}

% body of paper here - Use proper section commands
% References should be done using the \cite, \ref, and \label commands
% Put \label in argument of \section for cross-referencing
%\section{\label{}}

%%%%%%%%%%%%%%%%%%%%%%%%%%%%%%%%%%
\section{Introduction}

Though the standard model (SM) is established by the discovery of the SM-like Higgs boson,
new physics beyond the SM are still required 
for solving several serious problems such as
a mechanism to produce the baryon asymmetry of the Universe (BAU),
a origin of tiny neutrino masses, 
and a candidate of the dark matter (DM), 

It is interesting to focus on the scenarios which solves these three problems
at around the TeV scale, {\it i.e.} the electroweak baryogenesis\cite{EWBG} for the mechanism for 
producing BAU, radiative seesaw scenarios for the origin of tiny neutrino masses, 
and introducing weak interacting massive particles as candidates of the DM. 
Many nice models in such a direction have been developed in literature.
In particular, the Aoki-Kanemura-Seto (AKS) model\cite{Aoki:2008av} is an attractive example 
which includes all the three mechanisms.

However, in the AKS model, it is known that a Landau pole appears at the scale much below the Planck scale.
It means that there should be a more fundamental theory above a cutoff scale.
In the following, we review a candidate of such a fundamental theory proposed in 
Refs.~\cite{  Kanemura:2014cka,Kanemura:2012uy}, 
which is based on a supersymmetric (SUSY) gauge theory with confinement.

\section{The model}
It is known that confinement occurs in the SU$(N_c)$ SUSY gauge theory with $N_f$ flavours when 
$N_f=N_c+1$ is satisfied\cite{Intriligator:1995au}. 
The simplest case is $N_c=2$ and $N_f=3$.
We utilize this simplest setup and propose a SUSY SU(2)$_H$ gauge theory\footnote{It's the 
	same setup as the minimal SUSY fat Higgs model\cite{Harnik:2003rs}. In the minimal SUSY fat Higgs model, 
	only $H_u$, $H_d$, and $N$ are made light by introducing additional fields. On the other hand, all the mesonic fields listed 
	in Table~\ref{FieldContent}-(II) play an important role in our model.}
In order to forbid tree level contributions to neutrino masses, 
an unbroken $Z_2$ symmetry is introduced to the model.
We also introduce a right-handed neutrino (RHN) superfield which has odd number
under the $Z_2$ symmetry. The assignment of the SM charge and the $Z_2$-parity 
on the SU(2)$_H$ doublets and the RHN is shown in Table~\ref{FieldContent}-(I).

In this framework, the SU(2)$_H$ gauge coupling becomes strong at a certain scale $\Lambda_H$, and 
the low energy effective theory below $\Lambda_H$ is described in terms of the fifteen mesonic 
fields listed in the Table~\ref{FieldContent}-(II), where the mesonic superfields 
are canonically normalized as $H_{ij}\simeq \frac{1}{4\pi \Lambda_H}T_iT_j (i\neq j)$.
%by using the Naive Dimensional Analysis.
The superpotential of the Higgs sector in the low energy effective theory can be 
written as 
\begin{align}
		W_{\text{eff}}=&
	{\lambda}
		N\left(H_uH_d+v_0^2\right)
		+
	{\lambda}
		N_{\Phi}\left(\Phi_u\Phi_d+v_{\Phi}^2\right)
		+
	{\lambda}
		N_{\Omega}\left(\Omega_+\Omega_--\zeta\eta+v_{\Omega}^2\right)
		\nonumber\\
		&+
		{\lambda}\left\{
		\zeta H_d\Phi_u
		+\eta H_u\Phi_d
		-\Omega_+H_d\Phi_d
		-\Omega_-H_u\Phi_u
		-NN_{\Phi}N_{\Omega}
\right\}\;.
\end{align}
%By the Naive Dimensional Analysis, 
It is naively expected that $\lambda \simeq 4\pi$ at the confinement 
scale $\Lambda_H$.
The relevant part of the soft SUSY breaking Lagrangian is given by 
\begin{align}
	\mathcal{L}_H=&
	-m_{H_u}^2H_u^{\dagger}H_u
	-m_{H_d}^2H_d^{\dagger}H_d
	-m_{\Phi_u}^2\Phi_u^{\dagger}\Phi_u
	-m_{\Phi_d}^2\Phi_d^{\dagger}\Phi_d
	-m_N^2 N^*N
	-m_{N_{\Phi}}^2 N_{\Phi}^*N_{\Phi}
	-m_{N_{\Omega}}^2 N_{\Omega}^*N_{\Omega}
	\nonumber\\
	&
	-m_{\Omega_+}^2\Omega_+^*\Omega_+
	-m_{\Omega_-}^2\Omega_-^*\Omega_-
	-m_{\zeta}^2\zeta^*\zeta
	-m_{\eta}^2\eta^*\eta
	-\left\{
	m_{\zeta\eta}^2\eta^*\zeta +\frac{B_{\zeta}^2}{2}\zeta^2
	+\frac{B_{\eta}^2}{2}\eta^2 +\text{h.c.}
	\right\}
	\nonumber\\
	&
	-\left\{C\lambda v_0^2 N+C_{\Phi}\lambda v_{\Phi}^2 N_{\Phi} + C_{\Omega}\lambda v_{\Omega}^2 N_{\Omega}+\text{h.c.}\right\}
	-\left\{ B\mu H_uH_d + B_{\Phi}\mu_{\Phi}\Phi_u\Phi_d +B_{\Omega}\mu_{\Omega}(\Omega_+\Omega_-+\zeta\eta)+\text{h.c.}\right\}
	\nonumber\\
	&
	-\lambda\bigl\{ 
		A_NH_uH_dN
		+A_{N_{\Phi}}\Phi_u\Phi_dN_{\Phi}
		+A_{N_{\Omega}}(\Omega_+\Omega_--\eta\zeta)N_{\Omega}
		+A_{\zeta}H_d\Phi_u\zeta 
		\nonumber\\
		&\phantom{-\lambda\bigl\{}
		+A_{\eta}H_u\Phi_d\eta
		+A_{\Omega_-}H_u\Phi_u\Omega_-
		+A_{\Omega_+}H_d\Phi_d\Omega_+
	+\text{h.c.}\bigr\}\;.
\end{align}
After the $Z_2$-even neutral fields $N$, $N_{\Phi}$ and $N_{\Omega}$ get vacuum expectation values (vev's),
the mass parameters $\mu=\lambda \langle N\rangle$, $\mu_{\Phi}=\lambda \langle N_{\Phi}\rangle$ and $\mu_{\Omega}=\lambda\langle N_{\Omega}\rangle$
are induced.

The Yukawa couplings and the Majorana mass term of the RHN are given by 
\begin{align}
	W_N=&y_N^i N_R^c L_i\Phi_u + h_N^i N_R^c E_i^c\Omega_- +\frac{M_R}{2}N_R^cN_R^c 
	+\frac{\kappa}{2}NN_R^cN_R^c\;.
	\label{eq:Wnu}
\end{align}

\begin{table}[t]
	\caption{(I) The charge assignment under the SM gauge group (SU(3)$_c\times $SU(2)$_L\times $U(1)$_Y$  and the $Z_2$ parity on the 
		SU(2)$_H$ doublets $T_i$ and the RHN $N_R^c$.
	(II) The field content of the extended Higgs sector or the low energy 
	effective theory of the SUSY SU(2)$_H$ model.
	\label{FieldContent}}
		\begin{center}
			\begin{tabular}{cc}
				(I)&(II)\\
				\begin{tabular}[t]{|c|c|c|c|c|c|}\hline
				Superfield&SU(2)$_H$&SU(3)$_C$&SU(2)$_L$&U(1)$_Y$&$Z_2$\\ \hline
				$\displaystyle
				\begin{pmatrix}
					T_1\\
					T_2
				\end{pmatrix}$& 2&1&2&0&$+1$\\ \hline
					$T_3$&2&1&1&$+1/2$&$+1$\\ \hline
					$T_4$&2&1&1&$-1/2$&$+1$\\ \hline
					$T_5$&2&1&1&$+1/2$&$-1$\\ \hline
					$T_6$&2&1&1&$-1/2$&$-1$\\ \hline \hline
					$N_R^c$&1&1&1&0&$-1$\\ \hline
			\end{tabular}&
		\begin{tabular}[t]{|c|c|c|c|c|c|}\hline
		Superfield&SU(3)$_C$&SU(2)$_L$&U(1)$_Y$&$Z_2$\\ \hline
			$H_d\equiv\begin{pmatrix}
				H_{14}\\
				H_{24}\\
			\end{pmatrix}$
			&1&2&$-1/2$&$+1$\\ \hline
			$H_u\equiv\begin{pmatrix}
				H_{13}\\
				H_{23}\\
			\end{pmatrix}$
			&1&2&$+1/2$&$+1$\\ \hline
			$\Phi_d\equiv\begin{pmatrix}
				H_{15}\\
				H_{25}\\
			\end{pmatrix}$
			&1&2&$-1/2$&$-1$\\ \hline
			$\Phi_u\equiv\begin{pmatrix}
				H_{16}\\
				H_{26}\\
			\end{pmatrix}$
			&1&2&$+1/2$&$-1$\\ \hline
			$\Omega_-\equiv H_{46}$&1&1&$-1$&$-1$\\ \hline
			$\Omega_+\equiv H_{35}$&1&1&$+1$&$-1$\\ \hline
			$N\equiv H_{56},N_{\Phi}\equiv H_{34},N_{\Omega}=H_{12}$&1&1&$0$&$+1$\\ \hline
			$\zeta \equiv H_{36},\eta\equiv H_{45}$&1&1&$0$&$-1$\\ \hline
		\end{tabular}
	\end{tabular}
	\end{center}
\end{table}

\section{Benchmark points and its predictions}
In the low energy effective theory of the model, 
the first order electroweak phase transition (1stOPT) can be enhanced by the 
loop contributions of extra $Z_2$-odd scalar particles such as $\Phi_u$ and 
$\Omega_-$ strongly enough to satisfy the condition $\varphi_c/T_c>1$, which
is necessary for successful electroweak baryogenesis. 
Here, we focus only on the 1stOPT.
In order to reproduce the BAU, we should also require new CP violating phases.
We expect that we can introduce several new CP phases which contribute to 
the baryogenesis as in the case of MSSM\cite{CPV}.

\begin{figure}
	\begin{center}
		\begin{tabular}{ccc}
			\includegraphics[width=4cm]{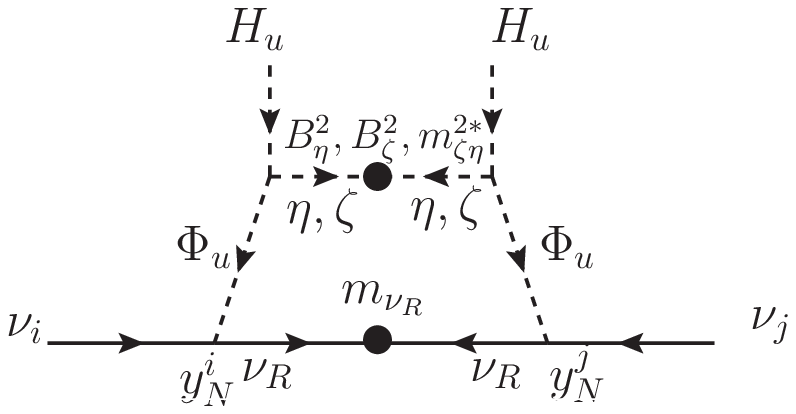}&
			\includegraphics[width=4cm]{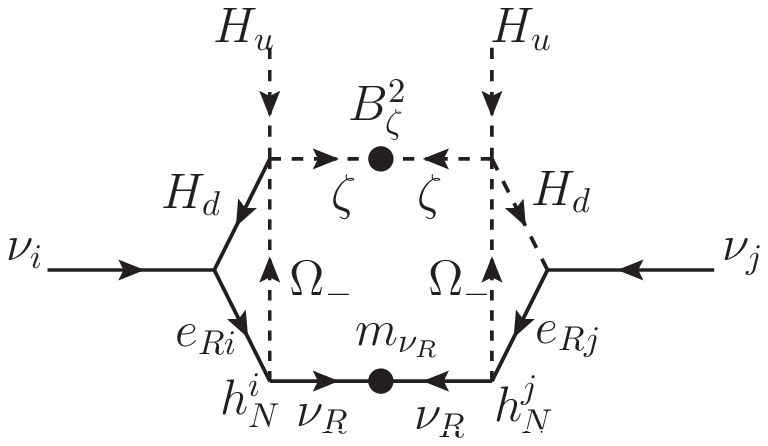}&
			\includegraphics[width=4cm]{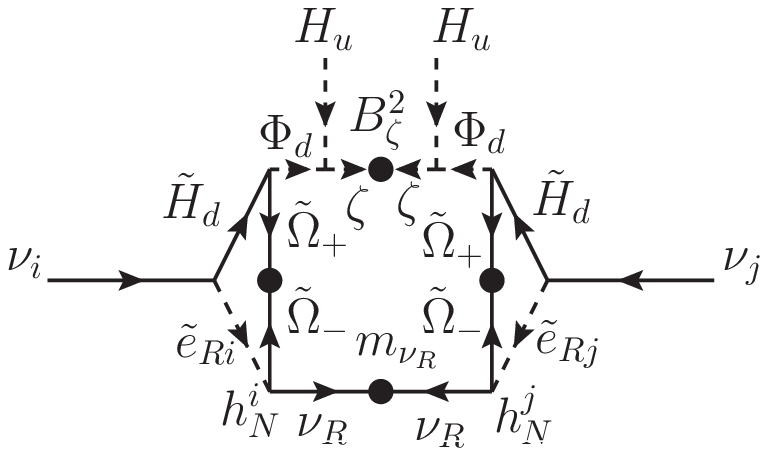}\\
			(I)&(II)&(III)\\
		\end{tabular}
	\end{center}
	\caption{(I) A one-loop diagram and (II) three-loop diagrams which contribute to the 
		neutrino mass matrix. The figures are taken from \cite{Kanemura:2014cka}} 
	\label{numassdiag}
\end{figure}

Tiny neutrino masses are generated at loop levels as shown in Fig.~\ref{numassdiag}. 
The one-loop diagrams are driven by the neutrino Yukawa coupling $y_N$ and the three-loop diagrams are
controlled by the coupling $h_N^i$. Because of this, two different mass squared differences are 
explained even if only one RHN is introduced.

Since both $Z_2$-parity and $R$-parity are unbroken in our model, 
there can be three kinds of the DM candidates, {\it i.e.} the lightest particles with 
the parity assignments of $(-,+)$, $(+,-)$, and $(-,-)$. If one of these three particle 
is heavier than the sum of the masses of the others,
the heaviest one decays and only the other two can be DM.

In the Table~\ref{Tab:Bench}, we list the definition of a benchmark scenario and 
its predictions, where the condition $\varphi_c/T_c>1$ is satisfied, the 
neutrino masses and the mixing angles given by neutrino oscillation data
can be reproduced, and the relic abundance of the DM can be explained 
with satisfying the constraints from the experiments such as 
LFV searches and the direct detection of the DM.
\begin{table}[ht]
	\caption{(i) The definition of our benchmark scenario, and 
		(ii) its predictions.
%	In the list, $\bar{m}_{\phi_i}^2=m_{\phi_i}^2+|\mu_i|^2$ are taken 
%	as input parameters, where $\mu_i=\mu_{\Phi}$ for $\phi_i=\Phi_u,\Phi_d$, and 
%	$\mu_i=\mu_{\Omega}$ for $\phi_i=\Omega_{+},\Omega_-,\zeta, \eta$.
	The tables are taken from Ref.~\cite{Kanemura:2014cka}.
	\label{Tab:Bench}}
	\begin{center}
		\begin{tabular}{c}
			(i) Input parameters for the benchmark scenario\\
		\begin{tabular}{|c|}\hline
			$\lambda$, $\tan\beta$, and $\mu$-terms \\ \hline
			$\lambda=1.8$ ($\Lambda_H=5$ TeV) \quad $\tan\beta=15$ \quad $\mu=250\;\text{GeV}$
			\quad $\mu_{\Phi}=550\;\text{GeV}$\quad 
			$\mu_{\Omega}=-550\;\text{GeV}$
			\\\hline\hline
			$Z_2$-even Higgs sector\\ \hline
			$m_h=126\;\text{GeV}$\quad $m_{H^{\pm}}=990$\;GeV \quad $m_N^2=(1050\;\text{GeV})^2$ \quad $A_N=2900\;\text{GeV}$
			\\ \hline\hline
			$Z_2$-odd Higgs sector \\ \hline
			$\bar{m}_{\Phi_u}^2=\bar{m}_{\Omega_-}^2=(175\;\text{GeV})^2$\quad 
			$\bar{m}_{\Phi_d}^2=\bar{m}_{\Omega_+}^2=\bar{m}_{\zeta}^2=(1500\;\text{GeV})^2$\quad
			$\bar{m}_{\eta}^2=(2000\;\text{GeV})^2$\\
			$B_{\Phi}=B_{\Omega}=A_{\zeta}=A_{\eta}=A_{\Omega^+}=A_{\Omega^-}=m_{\zeta\eta}^2=0$\quad
			$B_{\zeta}^2=(1400\;\text{GeV})^2$\quad $B_{\eta}^2=(700\;\text{GeV})^2$\\ \hline\hline
			RH neutrino and RH sneutrino sector\\ \hline
			$m_{\nu_R}=63\;\text{GeV}$\quad $m_{\tilde{\nu}_R}=65\;\text{GeV}$\quad $\kappa=0.9$\\
			$y_N=(3.28i, 6.70i, 1.72i)\times 10^{-6}$\quad 
			$h_N=(0,0.227,0.0204)$\\ \hline\hline
			Other SUSY SM parameters\\\hline
			$m_{\tilde{W}}=500\;\text{GeV}$\quad $m_{\tilde{q}}=m_{\tilde{\ell}}=5\;\text{TeV}$\\\hline
		\end{tabular}\\
		\\
		(ii) Predictions of the Benchmark points\\
		\begin{tabular}{|c|}\hline
			Non-decoupling effects\\\hline
			$\varphi_c/T_c=1.3$\quad $\lambda_{hhh}/\lambda_{hhh}|_{\text{SM}}=1.2$\quad 
			$\text{B}(h\to \gamma\gamma)/\text{B}(h\to \gamma\gamma)|_{\text{SM}}=0.78$\\ \hline\hline
			Neutrino masses and the mixing angles\\ \hline
			$(m_1, m_2, m_3)=(0, 0.0084\;\text{eV}, 0.0050\;\text{eV})$\quad
			$\sin^2\theta_{12}=0.32$\quad 
			$\sin^2\theta_{23}=0.50$\quad 
			$|\sin\theta_{13}|=0.14$\\ \hline\hline
			LFV processes\\ \hline
			$\text{B}(\mu\to e\gamma)=3.6\times 10^{-13}$\quad 
			$\text{B}(\mu\to eee)=5.6\times 10^{-16}$\\ \hline\hline
			Relic abundance of the DM\\ \hline
			$\Omega_{\nu_R}h^2=0.055$\quad 
			$\Omega_{\tilde{\nu}_R}h^2=0.065$\quad 
			$\Omega_{\text{DM}}h^2=\Omega_{\nu_R}h^2+\Omega_{\tilde{\nu}_R}h^2=0.12$\\ \hline\hline
			Spin-independent DM-proton scattering cross sections 
			\\ \hline
			$\sigma_{\nu_R}^{\text{SI}}=3.1\times 10^{-46}\,\text{cm}^{2}$\quad
			$\sigma_{\tilde{\nu}_R}^{\text{SI}}=7.7\times 10^{-47}\,\text{cm}^{2}$\quad
			$\sigma_{\text{DM}}^{\text{SI}}=1.1\times 10^{-46}\,\text{cm}^{2}$
			\\ \hline
		\end{tabular}
	\end{tabular}
	\end{center}
	\end{table}

\begin{figure}
\includegraphics[height=3.8cm]{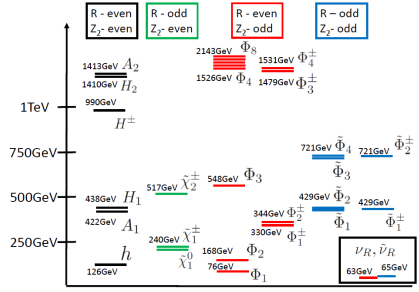}
\caption{The mass spectrum of the relevant particles in the bench mark scenario.
	The figure is taken from Ref.\cite{Kanemura:2014cka}.}
	\label{fig:masses}
\end{figure}

In Fig.~\ref{fig:masses}, we show the mass spectrum of the relevant particles in the benchmark scenario given in 
Table~\ref{Tab:Bench}
In this scenario, the $Z_2$-even sector is similar to the nMSSM which can be distinguished from the MSSM by 
the spectrum of extra Higgs bosons. For example, the mass splitting between the charged Higgs boson and 
the heavy Higgs bosons is caused by the large mixing between doublet fields and a singlet field, 
which is necessary in order to reproduce the relic abundance of the DM.

In the benchmark scenario, $\varphi_c/T_c$ is enhanced by the loop effect of $\Phi_u$ and $\Omega_-$,
which can also significantly affect
the $h$-$\gamma$-$\gamma$ coupling and the triple Higgs boson coupling as shown in Table~\ref{Tab:HiggsFingerPrint}.
By precise measurement at future collider experiment such as ILC\cite{ILC} of such the Higgs boson couplings, 
our benchmark scenario can be distinguished from nMSSM.
\begin{table}
	\caption{The deviations in the coupling constants 
	from the SM values in the benchmark scenario.\label{Tab:HiggsFingerPrint}}
	\begin{center}
	\begin{tabular}{|c|c|c|c|c|c|c|c|} \hline
		Couplings&$hWW$&$hZZ$&$h\bar{u}u$&$h\bar{d}d$&$h\bar{\ell}\ell$
			&$h\gamma\gamma$&$hhh$\\ \hline
			$\kappa_{h\phi\phi}=g_{h\phi\phi}/g_{h\phi\phi}^{\text{SM}}$&
			$0.990$&$0.990$&$0.990$&$0.978$&$0.978$&$0.88$&$1.2$\\ \hline
		\end{tabular}
\end{center}
\end{table}
In addition, the direct search of inert doublet particles\cite{Aoki:2013lhm} and 
inert charged singlet searches\cite{Aoki:2010tf} at ILC can also provide 
a strong hint on the $Z_2$-odd sector of the scenario. 

\section{Summary}
We have attempted to propose a simple model to explain the three problems such 
as baryogenesis, tiny neutrino mass, and DM in its low energy effective theory
and we have succeeded to find such a model based on SUSY SU(2)$_H$ gauge theory with confinement.
We have introduced a benchmark scenario and we have discussed how to test it at future collider experiments.

%%%%%%%%%%%%%%%%%%%%%%%%%%%%%%%%%%

% If you have acknowledgments, this puts in the proper section head.
%\bigskip % extra skip inserted
%%%%%%%%%%%%%%%%%%%%%%%%%%%%%%%%%%
\begin{acknowledgments}
This work is supported in part by JSPS KAKENHI, Grant Numbers 23104011 and 24340046.
\end{acknowledgments}

\bigskip % extra skip inserted
% Create the reference section using BibTeX:
%\bibliography{basename of .bib file}

\end{document}